\documentstyle[12pt,epsfig]{article}
\begin{document}
\begin{center}
{N.A. Korkhmazyan, N.N. Korkhmazyan \\
\it{5 Khanjyan St., Armenian Pedagogical Institute, \\
Yerevan 375010, Armenia E-mail: yndanfiz@ysu.am}}
\end{center}
\vspace{3cm}
\begin{center}
{ON SYMMETRY OF ELEMENTARY PARTICLES}
\end{center}
\vspace{3cm}
\begin{center}
{Abstract}
\end{center}

The new quantum number $\sigma$ is introduced. It is shown that the
conservation of $\sigma$-number results in the conservation of 
difference between baryon and lepton numbers obtained earlier by Georgi 
and Glashow.
The conservation of $\sigma $-number also predicts that electron type
neutrino mass is exactly zero. The quark-lepton symmetry is discussed. 
It is shown that the nature of quark-lepton symmetry is reflected by the 
fact that elementary particles of the same generation are subject to the 
symmetry transformation represented by 4-group of diedr. It is also shown 
that colorless elementary particles are subject to the same symmetry
transformation. The new elementary particles (transbaryons) are predicted.

\section{ The law of conservation of $\sigma $-number.}

It is known that upper and lower quarks are positioned asymmetrically on the
''charge axis''. With the aim to bring in some symmetry to this asymmetric
disposition we will introduce the new additive quantum number $\sigma $
determined so as to result, in combination with quark electric charge $q$,
the charge of the respective lepton.

Then for $u$ and $d$ quarks we will have, respectively, $\sigma _u=1/3$ and $
\sigma _d=-2/3$. In general, $\sigma $-numbers for all quarks and
anti-quarks are determined by the following formula:\cite{Kor1}

\begin{equation}
\sigma =q-1/3,\,\,\,\,\,\,\,\,\,\,\,\tilde{\sigma }=\tilde{q}+1/3.
\label{1}
\end{equation}

Now the quarks (and leptons) are positioned symmetrically with respect 
to $\sigma =-q$ axis on $(q,\sigma )$ plane (see Fig.1).
This brings about an idea that together with the law of conservation of
electric charge there is realized the second law of conservation, the law of
conservation of $\sigma $-number. As we will see later, this idea results in
extremely valuable findings. In particular, the conservation of $\sigma $
-number results in conservation of the difference between baryon and lepton
numbers obtained earlier by Georgi and Glashow from $SU(5)$ symmetry.\cite
{Geor}

Let us first show that $\sigma $-numbers for baryons, mesons, leptons and
photons are determined by the following expressions:

\begin{equation}
\sigma _B=Q_B-1,\,\,\sigma _M=Q_M,\,\,\,\sigma _L=Q_L+1,\,\sigma _\gamma
=0,\,\tilde{\sigma }=-\,\sigma   \label{2}
\end{equation}
where $Q$ is an electric charge of a particle. Indeed, formula (\ref{2}) is
obvious for baryons and mesons due to their quark structure. Similarly, 
$\sigma $-numbers can be obtained for exotic adrons. Using (\ref{2}), we will
obtain, in particular, $\sigma (p)=0,\,\sigma (n^0)=-1,\,\,\sigma (\pi
^0)=0\,,\,\,\sigma (\tilde{\Lambda }^0)=1,\;\sigma (\sum^{-})=-2\,$ and
so on. Taking into account that $\sigma (\pi ^0)=0\,$, we will obtain 
$\sigma _\gamma =0\,$from $\pi ^0\Rightarrow 2\gamma \,$decay.

In order to find $\,\sigma _L$, we will give the same quantum number (by
analogy with the electric charge) to all charged leptons $(\,\sigma
_e=\,\sigma _\mu =\,\sigma _\tau )\;$and another quantum number to neutral
leptons $(\,\sigma _{\nu _e}=\,\sigma _{\nu _\mu }=\,\sigma _{\nu _\tau })$
. These numbers are related to each other by the formula $\,\sigma _{\nu
_e}=1+\sigma _e\;$that can be derived from $n^0\Rightarrow p+e^{-}+
\tilde{\nu }\,$decay. Then from $uu\Leftarrow x\Rightarrow 
e^{+}\tilde{d}\;$reaction where the same boson may decay into antilepton +
antiquark or quark pair \cite{Okun} we derive $\,\sigma _{e^{+}}=-\sigma
_{e^{-}}=0$ , and $\,\sigma _\nu =1$. As we can see, these results are in
accordance with expressions (\ref{2}). After we have derived formula 
(\ref{2}), we can check out that the conservation of $\sigma $-number 
takes place
for all the reactions observed so far \cite{Europ}. It does not prohibit
also possible proton decay through channels $p\Rightarrow e^{+}\pi
^0;e^{+}\pi ^{+}\pi ^{-}$ where conservation of baryon and lepton numbers is
violated.

The general formula for conservation of $\sigma $-number can be written as
follows:

\begin{equation}
\sum Q_B-N_B+\sum \tilde{Q}_B+\tilde{N}_B+\sum Q_M+\sum \tilde{Q}_M + 
\sum Q_L+N_L+\sum \tilde{Q}_L-\tilde{N}_L=const \end{equation}
where $N$ is the number of particles. Here, taking into account the
conservation of electric charge, we will finally obtain the desired result:

\begin{equation}
(N_B-\tilde{N}_B)-(N_L-\tilde{N}_L)=const  \label{4}
\end{equation}

It should be pointed out that the baryon number in formula (\ref{4}) could
be replaced by total number of quarks$\,N_k$ comprised of baryon and meson
quark numbers. Indeed, taking into account that for baryons and mesons 
$(N_k-\tilde{N}_k)=3(N_B-\tilde{N}_B)\,$and $(N_k-\tilde{N}_k)=0\;$and
using formula (\ref{4}), we will obtain the following relation between 
quarks and leptons

\begin{equation}
(N_k-\tilde{N}_k)-3(N_L-\tilde{N}_L)=const  \label{5}
\end{equation}

As we will see later, the law of conservation of $\sigma $-number predicts
that electron type neutrino mass is exactly zero. It should be mentioned in
this relation that two contradictory theories trying to resolve the problem
of neutrino mass have been developed so far by Dirak-Weyl in 1929 and
Majoran in1936. According to the former \cite{Hal} electron neutrino mass is
exactly zero and the conservation of lepton number takes place. According to
the latter \cite{Umed,Cern} the electron neutrino mass is not zero and
therefore the conservation of lepton number does not take place. The Majoran
theory allows neutrinoless double beta decay where two $d$-quarks decay
through channel $d\Rightarrow u+e^{-}\;$with violation of lepton number.

The contemporary theories basing on $SU(5)$ and $SO(10)$ symmetry result in
the same contradictory predictions \cite{Okun} ''The $SO(10)$ symmetry
allows occurrence of some phenomena prohibited by $SU(5)$ symmetry. In
particular, $SU(5)$ theory predicts conservation of difference $B-L$ while
conservation of baryon number $B$ and lepton number $L\,$does not take
place. In $SO(10)$ theory, the difference $B-L$ may not hold constant if
sufficient number of Higgs fields are involved''. Approximately the same
results are mentioned in the Gell-Mann report.\cite{Gell}

Thus, the main question relating to neutrino mass remains open. There 
have been a number of attempts for the past 60 years to find theoretical 
or experimental solution but the problem still exists.\cite{Klap,Ger} The
recent experiments in many scientific centers worldwide have been aimed at
observing of neutrinoless double beta decay and neutrino oscillations.
Currently several new projects are proposed and huge scientific potential
and funds mobilized to get a breakthrough on this fundamental problem.
Twenty-three scientific centers are involved in the project MINOS (USA).\cite
{Fermi} The underground experiment is aimed at neutrino beams detecting
after passing some 730km from Fermi-lab (Viskonsin) to Soudan (Minnesota).
It is expected that some of the detected neutrinos will have changed their
flavor due to neutrino oscillations. In Japanese project KEK neutrino beam
will pass some 230km from an accelerator to the detector 
Super-Kamiokande. \cite{Fermi} The new Heidelberg project GENIUS is 
aimed at increasing the
sensitivity of Majorana neutrino mass from the present level of at best
0.1ev down to 0.01 or even 0.001ev. \cite{Klap}

Summarizing, we can say that the newly proposed law of conservation of 
$\sigma $-number prohibits neutrinoless double beta decay and, in 
compliance with Dirac-Weyl and $SU(5)$ theories, predicts that electron 
type neutrino mass is exactly zero.

\section{Quark-lepton symmetry.}

Now we will focus on the problem of quark-lepton symmetry.\cite{Okun,Hal}
The essence of the problem is to find such a symmetry among the elementary
particles, which would explain why each quark-lepton generation is
necessarily comprised of the pair of leptons and pair of quarks. Let us
shift to the new system of coordinates $(q^{^{\prime }},\sigma ^{^{\prime
}}) $ with the point of origin in the center of a rectangle $e\nu _eud$ and
axes directed along the axes of symmetry (see Fig1.). The relation with an
original system of coordinates is given by the following expressions:

\begin{equation}
q^{^{\prime }}=\frac {1}{\sqrt{2}}q+\frac {1}{\sqrt{2}}\sigma
,\;\,\,\,\,\,\,\,\,\,\,\,\,\sigma ^{^{\prime }}=-\frac 
{1}{\sqrt{2}}q+\frac {1}{\sqrt{2}}\sigma \;-\frac {1}{3\sqrt{2}}  \label{6}
\end{equation}

The coordinates of the vertices are now as follows:

\begin{equation}
e\left( -\frac {1}{\sqrt{2}},\frac{\sqrt{2}}{3}\right) ,\;\,\nu _e\left( 
\frac {1}{\sqrt{2}},\frac{\sqrt{2}}{3}\right) ,\;\,u\left( \frac 
{1}{\sqrt{2}},-\frac{
\sqrt{2}}{3}\right) ,\,\,\,d\left( -\frac {1}{\sqrt{2}},-\frac{\sqrt{2}}{3}\right);  
\label{7}
\end{equation}

The group that represents the symmetry transformation for the above
mentioned rectangle, consists of the following elements:
$E$-identity transformation; turn around axis $z$ for angle $2\pi $ ;
$A$-turn around axis $\sigma ^{^{\prime }}$ for angle $\pi $;
$B$-turn around axis $q^{^{\prime }}$ for angle $\pi $;
$C$-turn around axis $z$ for angle $\pi $.

For these elements we will obtain the group multiplication table (see Table
1).

\epsfig{file=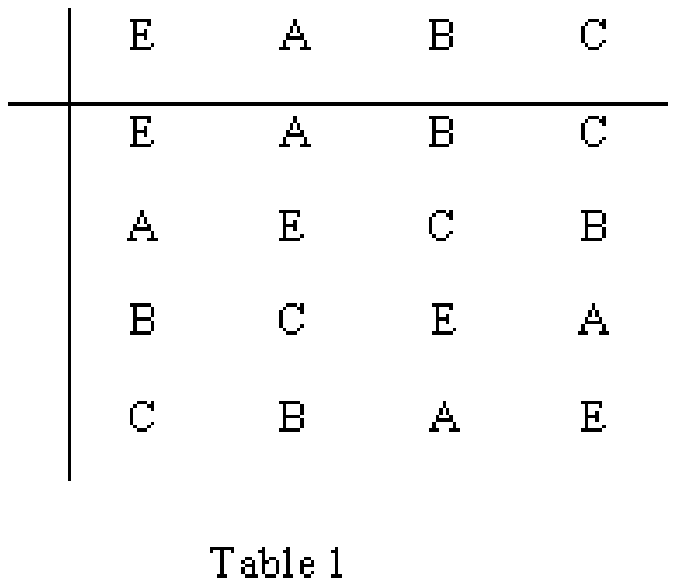}

Under multiplication we understand the subsequent execution of the
corresponding operations. The elements of the group have order 2 (except for
identity element ), since $\chi ^2=E\,$and $\chi ^{-1}=\chi \,$, where $\chi
\,$is an arbitrary element of the group. Thus, the totality of the 
elements $E,A,B,C$ makes up the Abelian group. The ''self-transformation'' of a
regular polygon is expressed by means of the following matrices \cite{Wig}

\begin{equation}
D_k=\left( 
\begin{array}{ll}
\,\,\,\,\cos \frac{2\pi k}{n}, & \sin \frac{2\pi k}{n} \\ 
-\sin \frac{2\pi k}{n}, & \cos \frac{2\pi k}{n}
\end{array}
\right) ,\,\,U_k=\left( 
\begin{array}{ll}
-\cos \frac{2\pi k}{n}, & \sin \frac{2\pi k}{n} \\ 
\,\,\,\,\,\sin \frac{2\pi k}{n}, & \cos \frac{2\pi k}{n}
\end{array}
\right).  \label{8}
\end{equation}

where $k=0,1,2,,n-1$. These $2n$-dimensional matrices make up the group of
order $2n$ known as the group of diedr. In case of n=2 we have the simplest
case of the group with elements

\begin{equation}
E=\left( 
\begin{array}{ll}
1 & 0 \\ 
0 & 1
\end{array}
\right) ,\,\,A=\left( 
\begin{array}{ll}
-1 & \,\,\,\,0 \\ 
\,\,\,0 & -1
\end{array}
\right) ,\,B=\left( 
\begin{array}{ll}
-1 & 0 \\ 
\,\,\,\,0 & 1
\end{array}
\right) ,\,C=\left( 
\begin{array}{ll}
1 & \,\,\,\,0 \\ 
0 & -1
\end{array}
\right).  \label{9}
\end{equation}

Taking into account that the group (\ref{9}) is isomorphic to the above 
mentioned group, we come to conclusion that (\ref{9}) is the matrix
representation of the group. Thus, if particles of the same generation are
located in the vertices of rectangle (\ref{7}) on the plane $(q^{^{\prime
}},\sigma ^{^{\prime }})$, this distribution is subject to the symmetry
transformation described by the group (\ref{9}). It is also easy to show
that if distribution of particles on the plane $(q^{^{\prime }},\sigma
^{^{\prime }})$ is subject to the symmetry transformation described by the
group (\ref{9}) and if coordinates of any arbitrary particle from (\ref{7})
coincide with one of the vertices, there should be three more particles (and
only three, without taking into account the color of the quarks) whose
coordinates coincide with the remaining vertices.

Summarizing this paragraph, we can say that the nature of the quark-lepton
symmetry could be described by $q\sigma $-symmetry. It is reflected by the
fact that particles of the same generation are subject to the symmetry
transformation represented by 4-group of diedr.\cite{Kor2}

\section{ Symmetry among colorless elementary particles.}

It is natural to expect that there is some symmetry among colorless
particles as well. We will consider all the baryons and mesons which can be
comprised of eighteen known quarks as well as the leptons. Each point on the
plane $(q,\sigma )$ is assumed to ''contain'' the whole family of the
particles (see fig.2). For example, point $(-1;0)$ contains, besides
electron, other charged leptons $\mu ^{-}\,$and $\tau ^{-}$. It is clear
from the picture that there is symmetry with respect to $\sigma ^{^{\prime
}}\,$axis. The number of particles in symmetrical points is identical. It is
easy to show that the same symmetry exists for multi-quark baryons as well.
For example, if baryons with quantum numbers $q=3,\sigma =-2,B=5$ consist of
eight upper and seven lower quarks, then symmetrical baryons with quantum
numbers $q=2,\sigma =-3,B=5$ consist of seven upper and eight lower quarks.
It is obvious that the number of colorless combinations is identical.

For baryons the same symmetry exists also with respect to other axis 
$q^{^{\prime }}$ which is determined by equation $\sigma =q-3$. It can be
proved by direct count of the particles in corresponding points (see fig.3).
The dashed lines parallel to $q^{^{\prime }}$ axis are determined by the
following formula:

\begin{equation}
\sigma =q-(B-L)  \label{10}
\end{equation}

It should be pointed out that formula (\ref{2}) can be represented by
formula (\ref{10}). For baryons, instead of (\ref{10}), we have $\sigma 
=q-B$. The point $q=3,\sigma =-3,B=6$contains only one baryon comprised 
of all the eighteen known quarks. The distribution of colorless particles shows
that, in order to have completed the symmetry, it is necessary to assume 
some particles not observed so far in the points marked by crosses. For 
example, there should be three particles in the point with coordinates 
$q=4,\sigma =-3\,\,$because there are three particles (leptons) in the 
symmetrical point. Now the distribution of colorless particles will be 
subject to the symmetry transformation described by 4-group of diedr 
(\ref{9}). The predicted particles should be located in five points 
symmetrically
positioned to those of leptons and mesons. These particles, like 
leptons, should not have quark structure because for them $B=6;7$. They 
are supposedly ten times heavier than baryons. We will call them transbaryons.
The number of transbaryons is less by one of the total number of leptons 
and mesons. If, by any selection rule, it is prohibited for some 
particles to be located in some arbitrary points, then, in compliance 
with above mentioned symmetry, the same number 
of particles should be missing out in symmetrically positioned points. 
But this does not pertain to transbaryons
because symmetrical to them leptons and mesons (most of them) have 
already been observed. The newly proposed symmetry for colorless 
particles does not depend on generation. Based on this symmetry alone, 
we cannot judge about forth generation.

Concluding this paragraph, we can say that within the frame of the 
proposed symmetry for colorless elementary particles the existence of 
transbaryons seems undisputable .\cite{Kor3}

\newpage
\epsfig{file=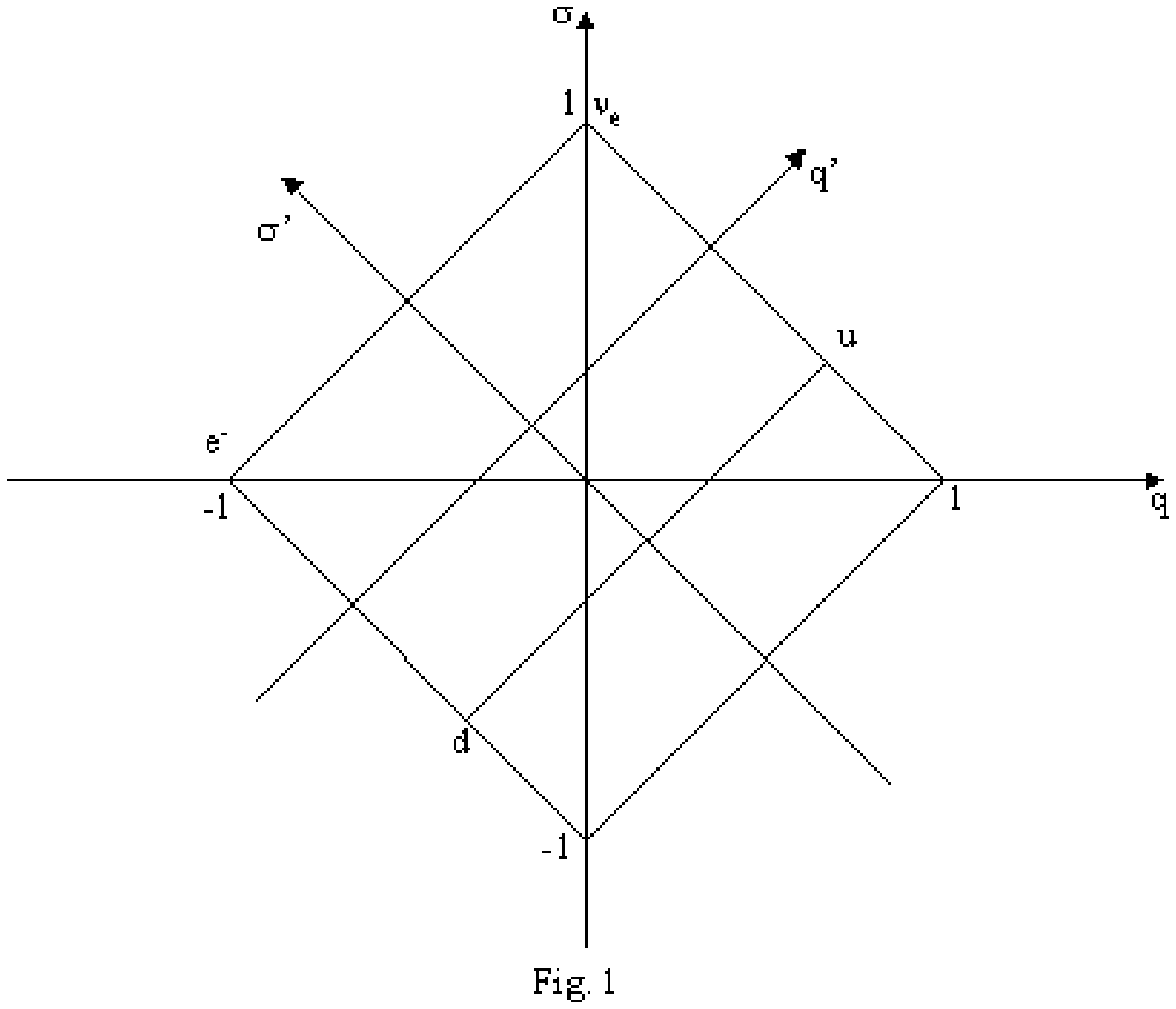,height=16cm, width=16cm}

Fig 1. Distribution of quarks and leptons of the first generation on 
$q,\sigma )$ plane.
\newpage
\epsfig{file=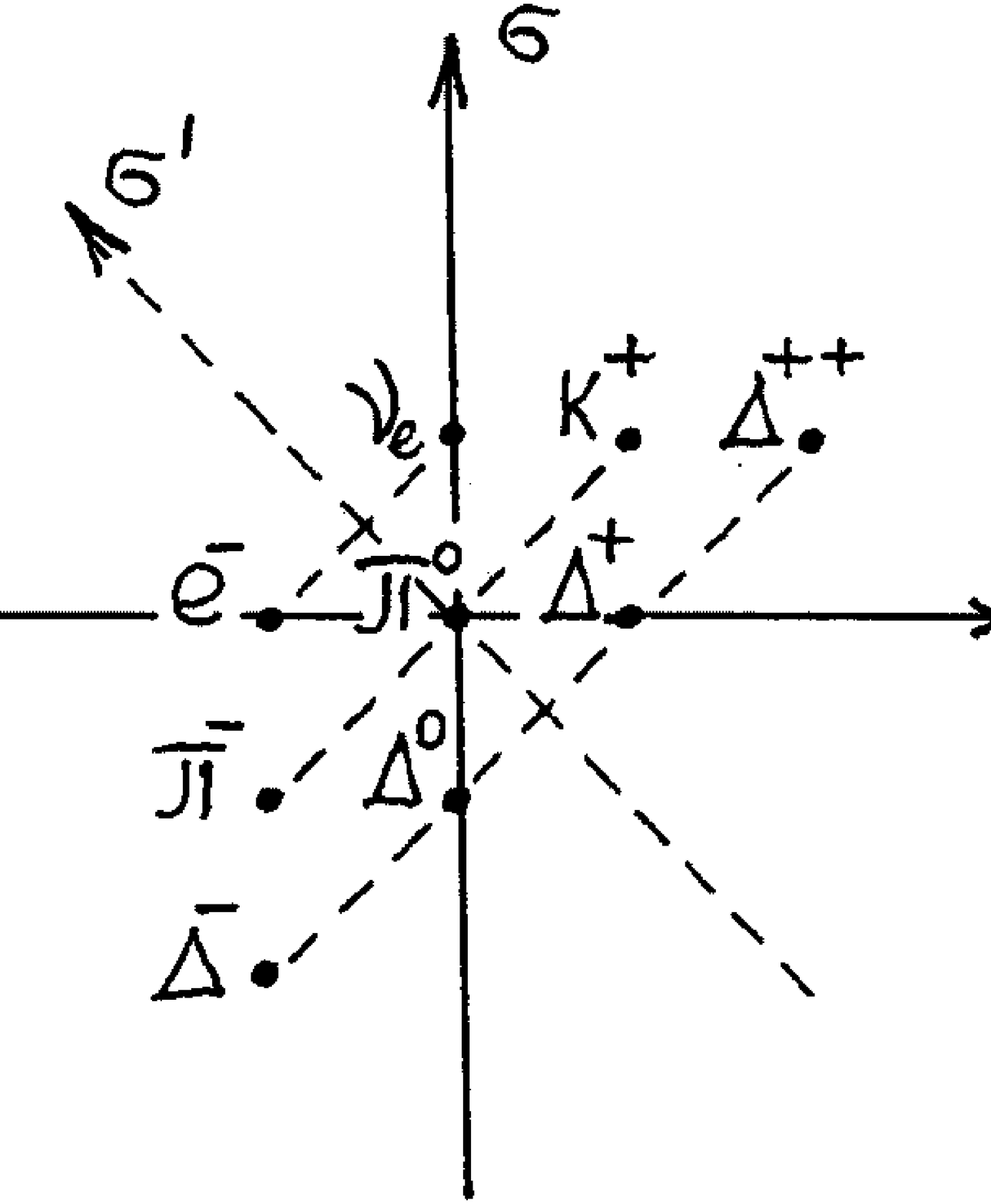,width=11cm, height=11cm}

Fig 2. Distribution of ordinary particles on $(q,\sigma )$ plane. The
distance between neighbouring points along axes $q$ and $\sigma $ is 
taken as a measurement unit.
\newpage
\epsfig{file=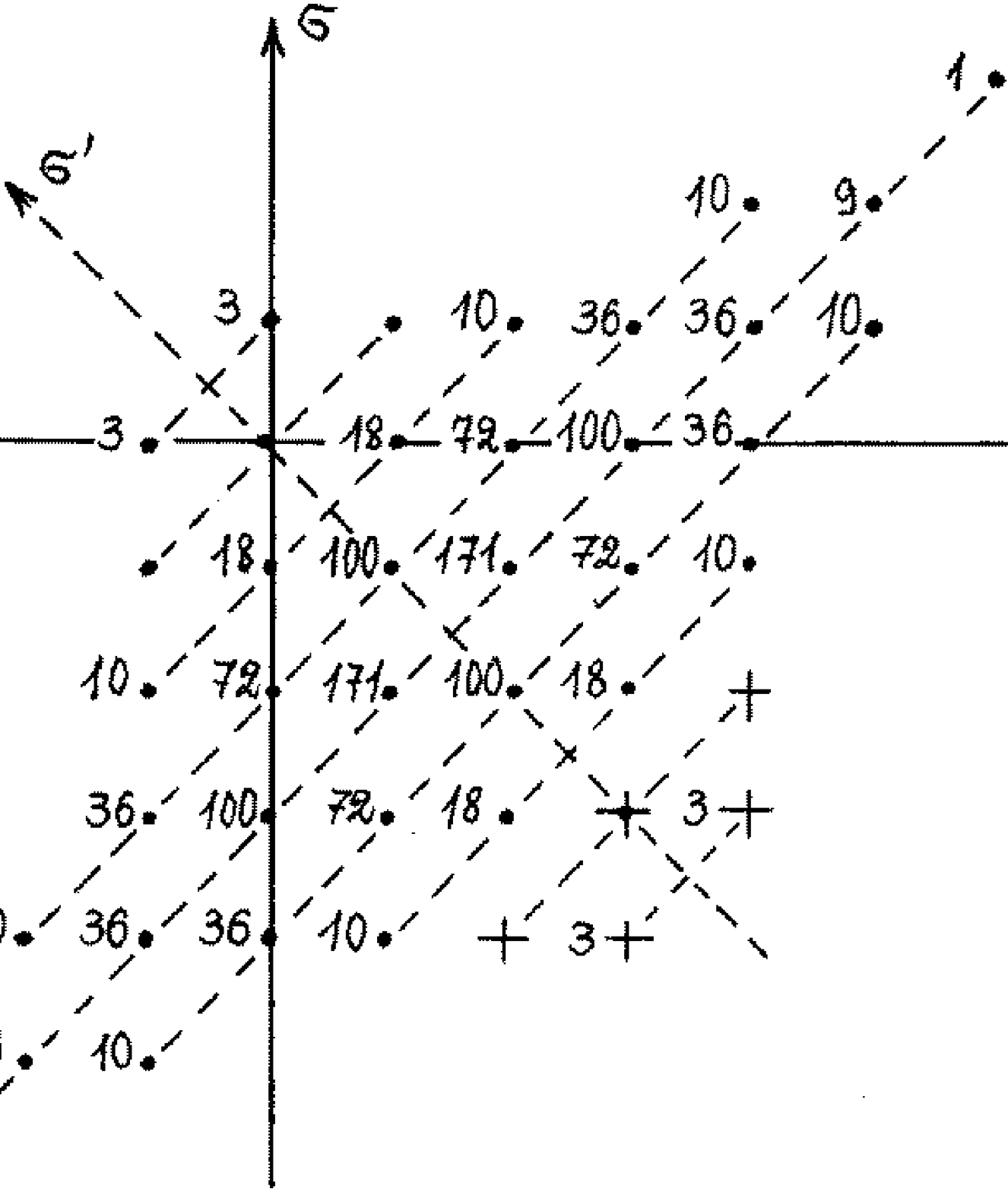,width=14cm,height=14cm}

Fig 3. Distribution of colorless particles on $(q,\sigma )$ plane.
Transbaryons are marked by crosses. The figures indicate the number of
particles in each point.
\end{document}